\documentclass[sigconf]{acmart}

\usepackage{colortbl}
\usepackage{enumitem}

\usepackage{caption}
\usepackage{subcaption}

\usepackage{tabularx}

\usepackage[multiple]{footmisc}

\usepackage{array}

\usepackage{booktabs}
\newcommand{\tabitem}{~~\llap{\textbullet}~~}

\AtBeginDocument{%
  \providecommand\BibTeX{{%
    \normalfont B\kern-0.5em{\scshape i\kern-0.25em b}\kern-0.8em\TeX}}}

\copyrightyear{2023}
\acmYear{2023}
\setcopyright{rightsretained}
\acmConference[WebSci '23]{15th ACM Web Science Conference 2023}{April
30-May 1, 2023}{Austin, TX, USA}
\acmBooktitle{15th ACM Web Science Conference 2023 (WebSci '23), April
30-May 1, 2023, Austin, TX, USA}
\acmDOI{10.1145/3578503.3583610}
\acmISBN{979-8-4007-0089-7/23/04}

\begin{document}

\title{Characterizing and Predicting Social Correction on Twitter}

\author{Yingchen Ma}
\affiliation{%
  \institution{Georgia Institute of Technology}
  \city{Atlanta}
  \state{Georgia}
  \country{USA}
}
\email{yma473@gatech.edu}

\author{Bing He}
\affiliation{%
  \institution{Georgia Institute of Technology}
  \city{Atlanta}
  \state{Georgia}
  \country{USA}
}
\email{bhe46@gatech.edu}

\author{Nathan Subrahmanian}
\affiliation{%
  \institution{Brandeis University}
  \city{Waltham}
  \state{Massachusetts}
  \country{USA}
}
\email{nsubrahmanian@brandeis.edu}

\author{Srijan Kumar}
\affiliation{%
  \institution{Georgia Institute of Technology}
  \city{Atlanta}
  \state{Georgia}
  \country{USA}
}
\email{srijan@gatech.edu}



\begin{abstract}

Online misinformation has been a serious threat to public health and society. Social media users are known to reply to misinformation posts with counter-misinformation messages, which have been shown to be effective in curbing the spread of misinformation. This is called social correction. However, the characteristics of tweets that attract social correction versus those that do not remain unknown. To close the gap, we focus on answering the following two research questions: (1) ``Given a tweet, will it be countered by other users?'', and (2) ``If yes, what will be the magnitude of countering it?''. 
This exploration will help develop mechanisms to guide users' misinformation correction efforts and to measure disparity across users who get corrected. 
In this work, we first create a novel dataset with 690,047 pairs of misinformation tweets and counter-misinformation replies. Then, stratified analysis of tweet linguistic and engagement features as well as tweet posters' user attributes are conducted to illustrate the factors that are significant in determining whether a tweet will get countered. Finally, predictive classifiers are created to predict the likelihood of a misinformation tweet to get countered and the degree to which that tweet will be countered. 
The code and data is accessible on \url{https://github.com/claws-lab/social-correction-twitter}.

\end{abstract}

\settopmatter{printfolios=true} 




\maketitle
\section{Introduction}
\begin{figure}[!t]
    \centering
    \includegraphics[width=0.35\textwidth]{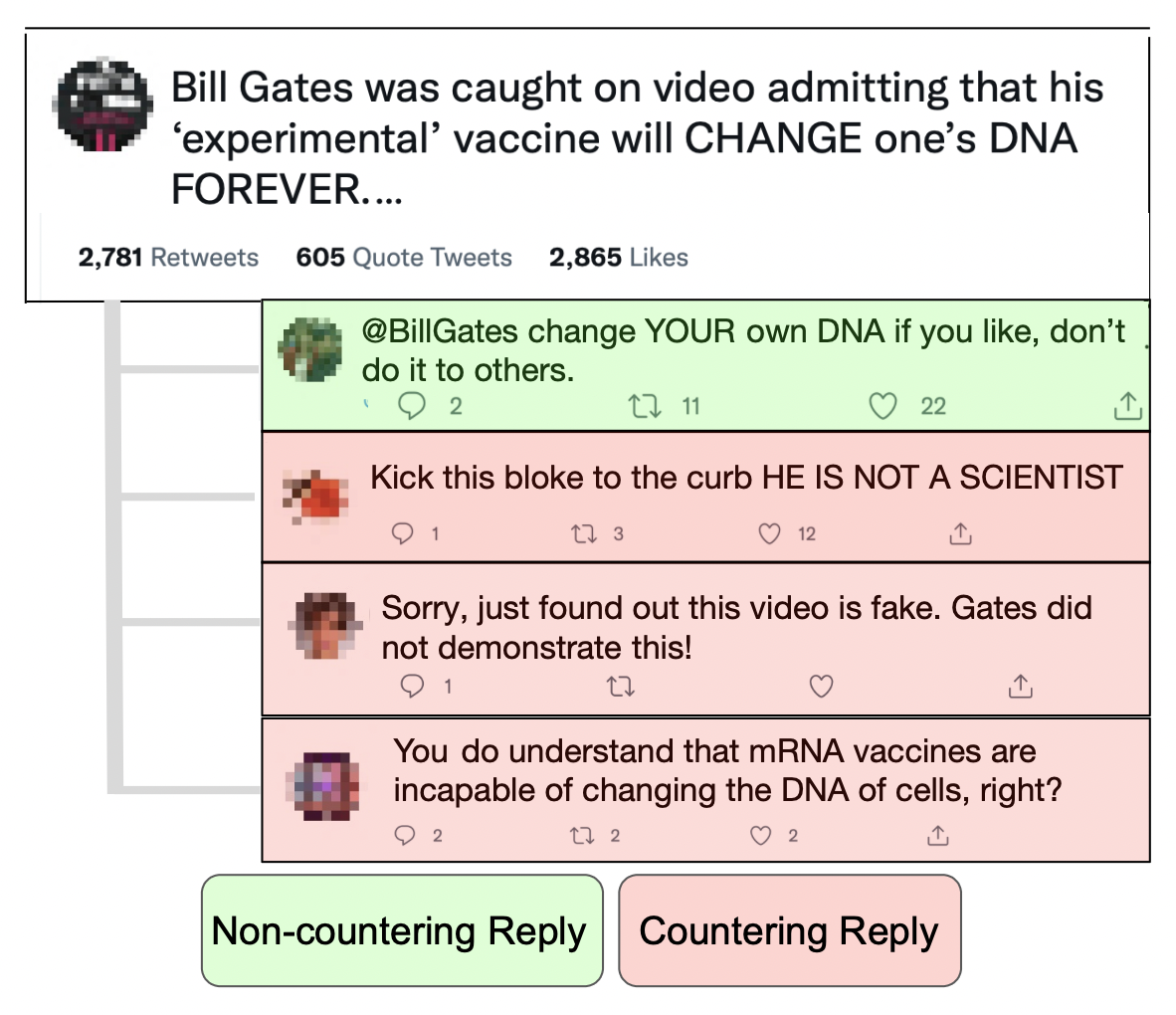}
    \caption{Examples of misinformation tweets and counter-misinformation replies. 
    }
    \vspace{-5mm}
    \label{fig:exampleTweet}
\end{figure}

Online misinformation leads to societal harm including diminishing trust in vaccines and health policies~\cite{pierri2022online,ball2020epic}, damaging the well-being of users consuming misinformation~\cite{kumar2018false,verma2022examining}, encouraging violence and harassment~\cite{arif2018acting,starbird2014rumors}, and posing a danger to democratic processes and elections~\cite{shin2017partisan,silverman2015lies,silverman2016analysis}. 
The problem has been exacerbated during the COVID-19 pandemic~\cite{micallef2020role, sharma2022covid}; particularly, COVID-19 vaccine misinformation including false claims that the vaccine causes infertility, contains microchips, and even changes one's DNA and genes has fueled vaccine hesitancy and reduced vaccine uptake~\cite{sharma2022covid}. Therefore, it is crucial to restrain the spread of online misinformation~\cite{micallef2020role, lewandowsky2012misinformation}.
In this work, we use a broad definition of misinformation which contains rumors, falsehoods, inaccuracies, decontextualized truths, or misleading leaps of logic~\cite{kumar2018false,wu2019misinformation}.

To combat misinformation, various countermeasures have been developed~\cite{micallef2020role, miyazaki2022fake, vo2019learning}. Recent work has shown that ordinary users of online platforms play a crucial role in countering misinformation. According to the research study by~\citet{micallef2020role}, the vast majority (96\%) of online counter-misinformation responses are made by ordinary users, with the remainder being made by professionals such as fact-checkers and journalists. While fact-checking from these professionals has been widely used due to its prominent and measurable impact~\cite{micallef2020role, vo2019learning}, this process typically does not involve engaging with the actors spreading misinformation. Instead, the ordinary users' counter-misinformation efforts complement those from professional fact-checkers by directly engaging in countering conversations through making independent posts or direct replies to misinformation posts made by others~\cite{he2023reinforcement}.

Countering of misinformation messages via direct replies from ordinary users is called \textit{social correction}~\cite{bode2018see, kligler2022collective}.
One real example is shown in Figure~\ref{fig:exampleTweet}. 
Notably, social correction has been shown to be effective in curbing the spread of misinformation~\cite{colliander2019fake, wijenayake2020effect}, as well as doing so without causing increases in misperception~\cite{guess2020does,swire2020searching,wood2019elusive}. While certainly not a panacea for convincing people to reconsider potentially misinformative beliefs, they are most effective at reducing the misperceptions of those who may consume it~\cite{bode2018see,Bode2021,colliander2019fake,seo2022if}.

However, little is known about the characteristics of misinformation tweets that attract social correction. 
Developing this understanding has several advantages: 
(1) first, it can help identify inequities in misinformation correction. For example, comparison of correction across users or communities (e.g., political ideologies) can reveal whether certain user types/communities are less likely to be self-correcting, e.g., communities where users correct misinformation when they see it. 
Identifying these disparities is the first step towards addressing them by redirecting resources towards entities that require external attention to curb misinformation; 
(2) second, if certain misinformation content is less likely to be socially corrected, targeted efforts can be directed toward countering them. 
Such instances can be escalated and prioritized for interventions by professionals or social media platforms;
(3) third, if certain misinformation content is likely to be socially corrected, then additional participants can be encouraged to provide reinforcements.

Despite these promising benefits, characterizing and predicting social correction is non-trivial due to several challenges. 
First, existing datasets do not contain conversation-style narratives with paired misinformation posts and counter-replies.
Second, existing works (including~\citet{miyazaki2022fake}) do not analyze counter-replies to misinformation in a stratified manner where tweets with different numbers of replies are considered separately. This fine-grained analysis is necessary since comparing across or aggregating statistics across tweets that have drastically different numbers of (counter-)replies can skew the findings~\cite{jeni2013facing, higgins2008meta}.

In this work, we seek to characterize and predict counter-replies to misinformation.
The contributions can be summarized as follows:
\begin{itemize}
    \item We curate a novel large-scale dataset that contains 1,523,849 misinformation tweets and 690,047 counter-misinformation replies, along with a hand-annotated dataset of misinformation tweets and counter-replies. 
    \item We perform a stratified, fine-grained analysis of the linguistic, engagement, and poster-level characteristics of misinformation tweets that get countered versus those that do not. Our analysis reveals several features of tweets that attract social correction, such as anger and impoliteness.
    \item We create two counter-reply prediction models to identify whether a misinformation tweet will be countered or not, and if so, to what degree (i.e. low or high), based on its linguistic, engagement, and poster features. We achieve promising predictive performance with both of these models, with best F-1 scores of 0.860 and 0.801, respectively.
\end{itemize}
The code and data is accessible on \url{https://github.com/claws-lab/social-correction-twitter}.

\section{Related Works}

\subsection{Social Correction on Social Media Platforms}

Misinformation widely spreads on social media platforms, which has caused detrimental effects on society~\cite{starbird2014rumors, arif2018acting, colliander2019fake}, including harassment and personal attacks~\cite{micallef2020role}. To combat misinformation, users actively employ various strategies~\cite{Mu2022IdentifyingAC}, including replying to and commenting on misinformation~\cite{vo2019learning, miyazaki2022fake, kligler2022collective}.
This debunking behavior can broadly reduce the misinformed beliefs of the author and the audience who see the misinformation~\cite{chan2017debunking, bode2018see}. Notably, current research works have shown the promising impact of debunking~\cite{chan2017debunking, bode2018see} in both curbing the perception of misinformation and reducing the belief of false information~\cite{chan2017debunking}. 
In this work, we deep-dive into this misinformation-countering behavior by looking at both the misinformation posts and the counter-misinformation replies to these posts. Since user response information can indicate the textual properties of misinformation posts that are highly likely to get countered, our work sheds light on better understanding of misinformation-countering behavior, especially, understanding the misinformation tweets that get countered. 

\subsection{Analysis of Counter-misinformation}\label{sec:related_work_analysis}
Due to the significance of counter-misinformation messages in curbing misinformation, much research has been focused on analyzing and understanding counter-misinformation~\cite{micallef2020role,vo2019learning}.

One type of work is to analyze and compare misinformation and counter-misinformation messages~\cite{micallef2020role,vo2019learning}. For instance, \citet{micallef2020role} first created a textual classifier to classify tweets into misinformation, counter-misinformation, and irrelevant groups, and then analyzed the tweets in each group. Interestingly, they find that a surge in misinformation tweets results in a corresponding  increase in tweets that reject such misinformation.
\citet{vo2019learning} first identified fact-checking replies by checking whether a reply contains a fact-checking URL toward two trustworthy fact-checking websites (i.e., Snopes.com and Politifact.com). Then, they retrieved the corresponding misinformation tweet toward which the fact-checking post replies to, and use them to construct pairs of misinformation posts and fact-checking replies for fact-checking content analysis and reply generation.

Meanwhile, ~\citet{miyazaki2022fake} curated a large-scale dataset containing pairs of misinformation tweets and debunking replies, by first crawling COVID-19 related misinformation tweets from existing research~\cite{cui2020coaid, kim2021fibvid, hossain2020covidlies, memon2020characterizing, shahi2021exploratory} and then recruiting crowd-sourcing workers via Amazon Mechanical Turk to annotate responses to these tweets as being debunking or not. They then perform analysis to illustrate who counters misinformation and how they do so.
However, contrary to this work, 
we conduct an in-depth \textit{stratified} analysis of the replies to examine which features matter during the countering. Stratification helps to compare similar tweets by controlling for the number of replies it receives. Furthermore, we also conduct analysis of whether tweets get a high or low proportion of countering replies. Importantly, we also build two new tasks of predicting which misinformation posts will get countered and to what degree they get countered.
Our work complements the existing counter-misinformation studies. 

\subsection{Birdwatch (a.k.a. Community Note)}
Twitter launched Birdwatch (recently renamed to Community Note) to facilitate misinformation detection by ordinary users. On the platform, users can report suspicious and/or misleading tweets, as well as annotate tweets reported by others. Many have investigated this kind of countering~\cite{allen2022birds, mujumdar2021hawkeye} and derived different patterns among this collective countering. For instance, ~\citet{allen2022birds} looks at the impact of partisanship during the crowds' annotation by analyzing existing data from the Birdwatch/Community Note platform; they find its users are more likely to (1) give negative annotations of tweets from counter-partisans, and (2) rate annotations from counter-partisans as unhelpful. Though Birdwatch/Community Note enables community-based detection of misinformation, it does not provide a way for users to counter misinformation. Notably, users provide inputs within the Birdwatch ecosystem only, which is restricted and does not reflect the larger dynamics of information flow on Twitter. 
Recent research has also shown that Birdwatch can be manipulated by motivated bad actors~\cite{mujumdar2021hawkeye}.
Therefore, we focus on the misinformation that spreads on Twitter and is countered by ordinary users for a more complete and comprehensive study.

\section{Dataset}

In this section, we describe 
the definition of the problem, as well as the corresponding dataset curation.

\subsection{Definitions}\label{sec:def}\hfill

\textbf{Misinformation}: We employ a broad definition of misinformation which includes falsehoods, inaccuracies, rumors, or misleading leaps of logic~\cite{wu2019misinformation}. Building on the existing work~\cite{hayawi2022anti}, we focus on misinformation related to the COVID-19 vaccine due to its broad impact around the world during the COVID-19 pandemic. Practically, the misinformative claims include ``the vaccine alters DNA'', ``the vaccine causes infertility'', ``the vaccine contains dangerous toxins'', and ``the vaccine contains tracking devices''; these topics are popular and widely studied by existing research works~\cite{hayawi2022anti, abbasi2022widespread}.

\textbf{Counter-reply}: Motivated by existing research works on analyzing replies that show disbelief toward misinformation~\cite{jiang2020modeling} or fact-check misinformation~\cite{pereira2022characterizing}, a direct response to a misinformation post $m$ is considered a ``countering'' reply
if it makes an attempt to explicitly or implicitly debunk or counter the misinformation tweet $m$. 
Otherwise, the reply is considered as non-countering. Practically, given a reply $r$, it is a:
\begin{itemize}
    \item \textbf{Countering reply:} Motivated by existing research works on identifying and analyzing text that is countering, debunking, disbelieving,  or disagreeing with misinformation~\cite{micallef2020role, jiang2020modeling, hossain2020covidlies}, a countering reply is a reply that explicitly or implicitly refutes the misinformation post (``this is misinformation''), points out the falsehood (``the COVID-19 vaccine does not change DNA''), insults the tweet poster (``you are born to lie''), or questions the misinformation (``Is there any reference I can check?'').
    \item \textbf{Non-countering reply:} Instead of countering, a non-countering reply supports, is in favor of, comments, repeats misinformation, etc., such as ``This is not the vaccine but the gene therapy'', ``Yes, I agree with you'', or ``It makes sense''. 
\end{itemize}
A post $m$ is considered to be countered if it receives at least one counter-reply. 
Meanwhile, given that different misinformation tweets have various numbers of replies, to have a normalized measure of the magnitude of which a misinformation tweet gets countered, we define the proportion of counter-replies to total replies, denoted as $p(m)$.

\subsection{Task Objective}
We consider the set $\mathcal{M}$ of misinformation posts about the COVID-19 vaccine. 
Each misinformation post $m \in \mathcal{M}$ has a set of $n$ replies $r = [r_1, ..., r_n]$ posted in direct response to $m$. 
Our final goal is to build a classifier $\mathcal{F}$ such that it can output a binary label $\mathcal{F}(m)$, which indicates whether the misinformation post will be countered or not, i.e., whether it will receive at least one counter-reply.

\subsection{Dataset Curation}

\subsubsection{Misinformation Tweet Collection and Classification}

We utilize the Anti-Vax dataset from~\citet{HAYAWI202223}, 
a large-scale dataset of tweets related to the topic of COVID-19 vaccines, in order to identify misinformation tweets for our study. These tweets range eight months from December 1, 2020 to July 31, 2021, which was the relevant period covering a substantial part of the time from when the vaccines were approved by the FDA in December 2020~\cite{sharma2022covid}.
Also during this period, many uncertainties and misinformation about COVID-19 vaccines were spreading on social media~\cite{muric2021covid, sharma2022covid, HAYAWI202223}. 
The original dataset consists of approximately 15.4 million  tweets collected from the Twitter API~\cite{HAYAWI202223}, each containing at least one of the following COVID-19 vaccine relevant keywords: \{`vaccine', `pfizer', `moderna', `astrazeneca', `sputnik', `sinopharm'\}. Only original tweets were considered, i.e., retweet, reply, or quote tweets were removed. We utilized the Twitter API to retrieve the tweet text, user ID of the tweet author, datetime, conversation ID, reply settings, and tweet engagement metrics (like, retweet, quote, and reply counts). In total, we were able to retrieve 14,123,209 tweets from the original dataset while the remaining 1.3 million tweets were unavailable due to the deletion by the users or the Twitter platform.

Following the definition of misinformation in Section~\ref{sec:def} and the current approach of identifying COVID-19 vaccine related misinformation tweets~\cite{HAYAWI202223}, we first get the annotated misinformation tweets from~\citet{HAYAWI202223}, train a text classifier to determine if a tweet is misinformation or not, and classify all non-annotated tweets. Specifically, we first crawl and get 4,836 annotated misinformation and 8,596 annotated non-misinformation tweets from~\citet{HAYAWI202223}. Next, we build a text classifier using BERT~\cite{devlin2019bert}. This classifier has a promising performance in precision, recall, and F-1 scores of $0.972$, $0.979$, and $0.975$, respectively. This performance is comparable to the reported one in the original paper by ~\citet{HAYAWI202223} (i.e., the precision, recall, and F-1 scores of $0.97$, $0.98$, and $0.98$). 
The classifier has high performance as per the metrics and thus can be used for downstream classification tasks. 

Finally, we use this misinformation classifier to identify misinformation tweets in the entire dataset, resulting in 1,523,849 misinformation tweets and 12,599,360 non-misinformation tweets. Since we only focus on replies to misinformation in this work, we only use misinformation tweets for downstream analyses.

Next, we perform filtering of the dataset. Since our work focuses on categorizing misinformation by the composition of their replies, we further discard misinformation tweets that have zero replies. 
In addition, we discard tweets where the poster has limited the set of users who can reply to their tweet, to ensure that all tweets in our dataset have an equal opportunity to be replied to. This information is obtained from the Twitter API.

Finally, our COVID-19 vaccine misinformation tweet dataset consists of 268,990 tweets where each tweet has at least one reply. This is the final set of misinformation tweets that we use. 

\subsubsection{Counter-misinformation Reply Collection and Classification}
\begin{figure}[h]
    \centering
    \includegraphics[width=0.4\textwidth, keepaspectratio]{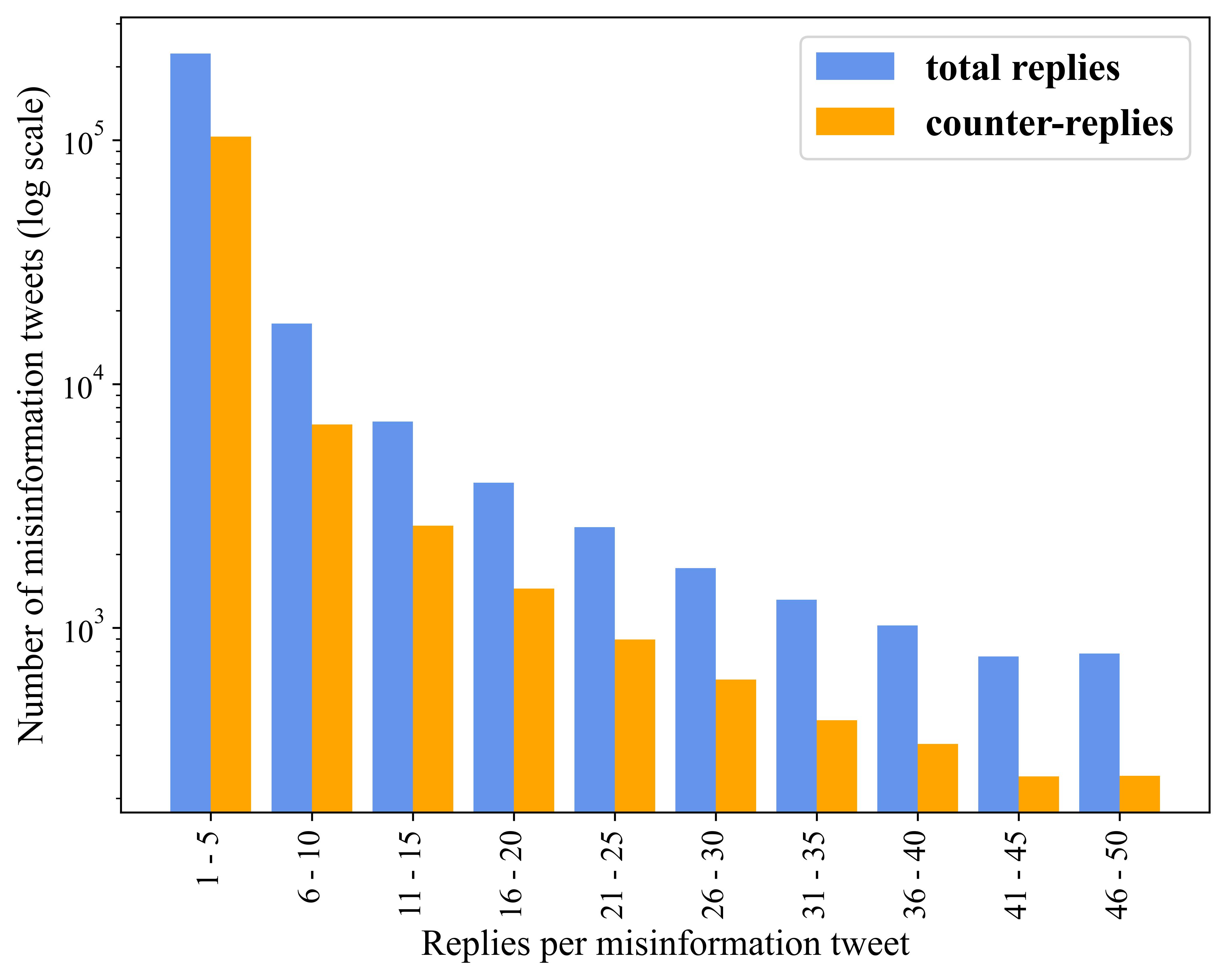}
    \caption{Distributions of the total number of replies (blue) and number of counter-replies (orange) per misinformation tweet, each presented on a log scale.}
    \label{fig:replies_per_tweet}
\end{figure}

For each tweet in our misinformation dataset, we use the Twitter API to crawl all direct replies to the original tweet. 
In total, we collected a total of 1,991,611 replies to the 268,990 tweets. One misinformation tweet has an average of approximately 7.4 replies. 
The distribution of the reply count per tweet is shown in Figure \ref{fig:replies_per_tweet} in blue.

\textbf{Building a Counter-reply Classifier:} Since it is of high cost to manually annotate all replies, in order to identify all the counter-replies (and non-counter-replies) from this set of 1.9 million replies, we train another text-based classifier to determine if a reply counters the tweet or not. 
Here, we call this a "counter-reply classifier".

Building on the existing works of the reply classification task~\cite{jiang2020modeling}, we first annotated replies and then built the classifier. Specifically, two students each first annotated 500 randomly-selected pairs of tweets and replies based on the textual contents into `Countering' or `Non-Countering' as per the definition provided in Section~\ref{sec:def}. 
This annotation resulted in an inter-rater agreement score of 0.7033 measured by percent agreement, resulting in 244 responses expressing countering while the remainder were non-countering. 
Then, after discussing the disagreements and creating the same annotation standard, each annotator labeled another 545 randomly selected pairs of tweets and replies. 
In total, we get 802 countering replies and 788 non-countering replies in our final annotated counter-reply dataset.

After getting the annotated replies, we utilize the Roberta-base lower-case architecture~\cite{liu2019roberta} as the classifier to which the input is the pair of tweets and replies. After the hyperparameter search across
batch size and learning rate, the classifier achieves a decent performance with a precision of 0.834, a recall of 0.819, and an F1-score of 0.822, which is sufficient for counter-reply classification on unlabeled replies.

Finally, we classify 690,047 (34.65\%) replies as counter-replies, and the remaining 1,301,564 (65.35\%) as non-counter-replies. The distribution of the counter-reply count per tweet is shown in Figure \ref{fig:replies_per_tweet} in orange. The average number of counter-replies that a misinformation tweet has is 2.57, and the average proportion of all replies of a misinformation tweet that are counter-replies is 0.271. 

\subsubsection{Misinformation Poster Attribute Collection}
For each misinformation tweet, we also collect information of the user who posted the misinformation tweet, which includes date and time of account creation, number of tweets posted, account verification, follower count, and following count. In total, information for 137,929 unique users was retrieved.

Additionally, we collected all the tweets that the user posted in the 7 days leading up to them posting the misinformation tweet; we refer to these tweets as ``pre-misinformation" tweets. Only original and quote tweets were retrieved; replies and other retweets were excluded. We pull the same set of attributes as in the misinformation tweet crawling. In total, we retrieved a total of 31,450,114 ``pre-misinformation'' tweets, with an average of 116.9 ``pre-misinformation'' tweets per misinformation tweet. Note that these numbers include duplicate tweets if the user had posted two misinformation tweets within 7 days of each other.

As a final step, we identify the subset of ``pre-misinformation" tweets that are related to the topic of COVID-19 vaccines, as well as those that are also misinformative. We define a ``pre-misinformation'' tweet belonging to that subset if it contains at least one of the aforementioned six keywords that were used to collect the original Anti-Vax dataset, namely \{`vaccine', `pfizer', `moderna', `astrazeneca', `sputnik', `sinopharm'\}.
In total, 1,781,161 (5.71\%) of the ``pre-misinformation'' tweets are labeled as being about COVID-19 vaccines. We then utilize the aforementioned misinformation classifier to identify COVID-19 vaccine misinformation within this subset of ``pre- misinformation'' tweets, of which 335,458 (18.83\%) were classified as misinformative.
\section{Characterization of Counter-reply}\label{sec:feature_analysis}

In this section, we analyze the properties of misinformation tweets with respect to the degree to which their misinformation gets countered. In order to do so, we identify the tweets that see a high proportion of their replies being counter-replies, and compare it to the group that see a low proportion of their replies being counter-replies.

To avoid skewing the results due to extreme data points, for this analysis, we do not consider tweets at the two extremes of the ``reply count'' distribution -- specifically, we remove tweets with fewer than three replies, as well as the top 2\% of tweets that have the greatest number of replies, following similar tweet filtering procedures in existing research works~\cite{yuan2016will, aleksandric2022twitter, alvarez2016topic}.
This is done to remove dataset noise related to low-engagement tweets, along with outliers associated with the highest engagement tweets. After this process, we are left with 74,663 misinformation tweets, with reply counts ranging from 3 to 52 (both inclusive).

\begin{figure}[h]
    \centering
    \includegraphics [width=0.4\textwidth, keepaspectratio]{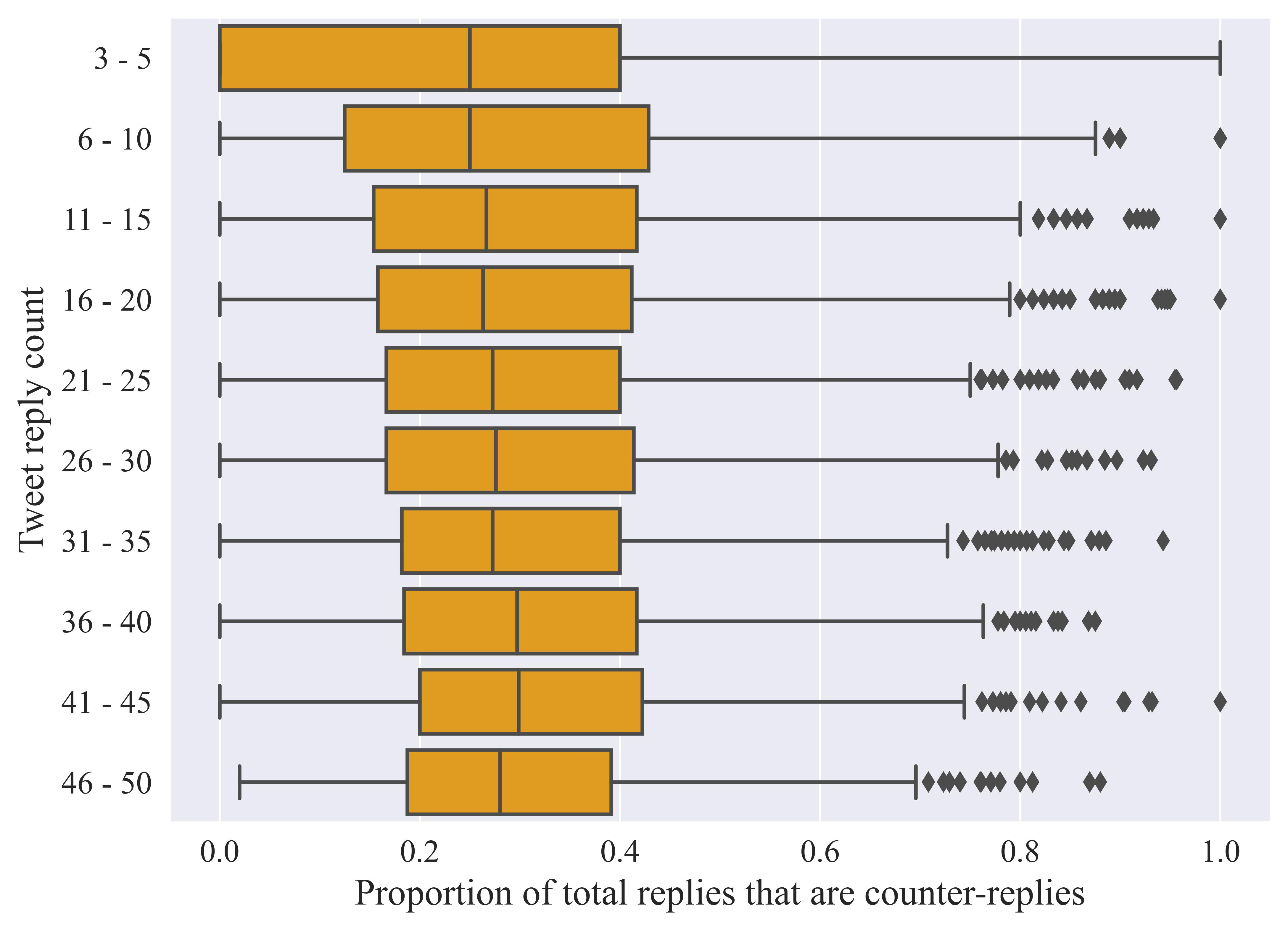}
    \caption{Distribution of proportion of counter-replies for each stratum. Each boxplot represents a stratum, displaying the minimum, maximum, quartiles, and (any) outliers.}
    \label{fig:disbelief_prop}
\end{figure}

\begin{figure}[h]
    \centering
    \includegraphics [width=0.4\textwidth, keepaspectratio]{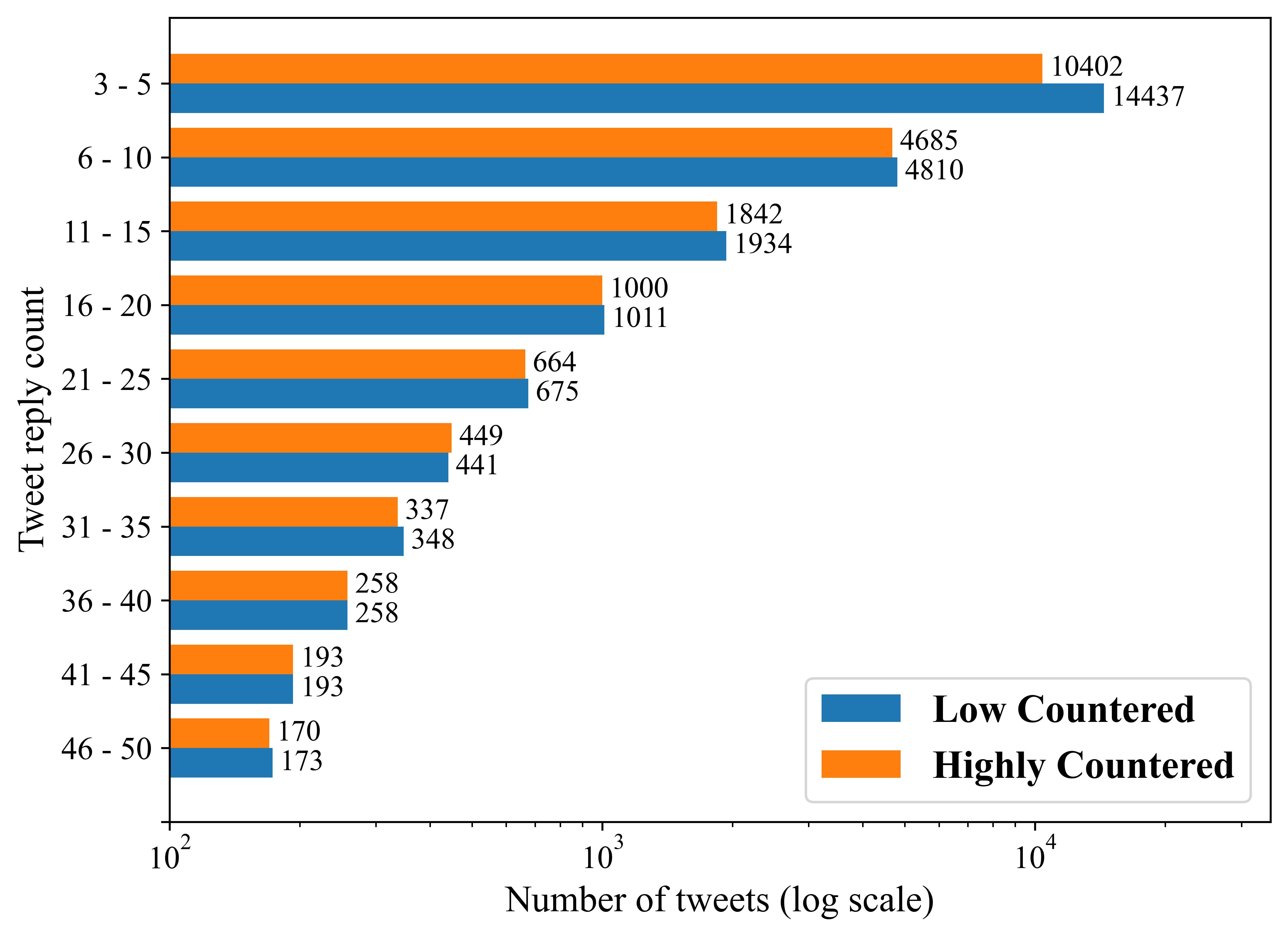}
    \caption{Number of tweets in each of the ``Low Countered'' (yellow) and ``Highly Countered'' (red) groups for each stratum, presented on a log scale.}
    \label{fig:low_vs_high}
\end{figure}

\subsection{Stratified Dataset Creation}\label{sec:feature_analysis_strat}

The linguistic, engagement, and user-level properties of tweets that get a low number of replies are different from those of tweets that receive many replies~\cite{matsumoto2019analysis, bruns2013towards, bruns2012quantitative}. Thus, to avoid conflating the factors that lead to receiving a high number of replies with the factors to receive counter-replies, we define and create several strata based on the number of replies that a misinformation tweet receives. 
Specifically, the strata are defined as follows: \textit{[3, 5], [6, 10], [11, 15], ..., [46, 50]}. 
Each stratum contains similar misinformation tweets that receive a similar number of replies, with some tweets that get countered and others that do not. We then compare these two groups within each stratum.
Figure \ref{fig:disbelief_prop} shows the distribution of counter-reply proportion within each stratum. We observe that, with the exception of tweets with a lower number of replies (that have more tweets with relatively fewer counter-replies), the distribution is similar across reply counts.

Within each stratum, we assign tweets to a ``Highly Countered'' group if its counter-reply proportion is in the top quartile (also within that stratum), a ``Low Countered'' group if its counter-reply proportion is in the bottom quartile (within that stratum), or discard it if it does not fall into either of the two groups. 
Figure \ref{fig:low_vs_high} shows the distribution of the tweets in the two relevant categories.

Within each stratum, we compare misinformation tweets between the two groups. We identify three types of attributes to perform this comparison along:
\begin{enumerate}
    \item \textbf{Tweet linguistic attributes}, to analyze the degree to which the tweet falls into meaningful personal, psychological, topical, emotion, and other content-related categories.
    \item \textbf{Tweet engagement attributes}, to analyze how and how much the tweet is interacted with among online users.
    \item \textbf{Tweet poster attributes}, to analyze the behavior, popularity, and status of the user behind the tweet.
\end{enumerate}
Table~\ref{tab:attr_list} displays the full list of attributes we study within each of these categories. We present results in the following subsections.

\begin{table*}[ht!]
    \centering
    \setlength\extrarowheight{-5pt}
    \begin{tabular}{p{0.15\linewidth} | p{0.8\linewidth}}
     \toprule
      Attribute type & List of attributes  \\
        \midrule
      Tweet linguistic & \tabitem number of words in the tweet*** \\
      & \tabitem VADER~\cite{Hutto_Gilbert_2014} positive sentiment, negative sentiment***, and compound sentiment*** of the tweet \\
      & \tabitem Politeness*** and impoliteness*** scores of the tweet, computed as the total number of linguistic strategy instances in the tweet positively and negatively correlated (respectively) with politeness as proposed by~\cite{DanescuNiculescuMizil2013ACA}. \\
      & \tabitem For each of the 65 (47*** + 18) dimensions of the LIWC~\cite{pennebaker2001linguistic} 2007 lexicon, the number of words for that dimension. \\
      \midrule
      Tweet engagement & \tabitem number of replies***, likes***, retweets*** (RTs), and quote tweets (QTs)*** \\
      & \tabitem number of likes, retweets (RTs), and quote tweets (QTs)***, each divided by the number of replies \\
      \midrule
      Tweet poster & \tabitem number of followers, number of users following***, whether the user is verified (1) or not (0)*** \\
      & \tabitem Total number of tweets the user has posted since account creation*** \\
      & \tabitem In the week (7 days) leading up to the misinformation tweet: the average \# of tweets posted per day***, the median count of likes*** and retweets*** received on their tweets, the number of tweets the user posted about COVID-19 vaccines***, and the proportion of COVID-19 vaccine tweets that are misinformation. \\
      \bottomrule
    \end{tabular}
    \caption{List of linguistic, engagement, and poster attributes considered for the analysis in Section~\ref{sec:feature_analysis}. A set of three asterisks(***) next to the attribute indicates a statistical test result of p < 0.001.\protect\footnotemark  This subset of statistically significant attributes is considered for the predictive tasks in Section~\ref{sec:prediction}.}
    \label{tab:attr_list}
\end{table*}

\begin{figure*}[h]
    \centering
    \begin{subfigure}[t]{0.24\textwidth}
        \raisebox{-\height}{\includegraphics[width=\textwidth]{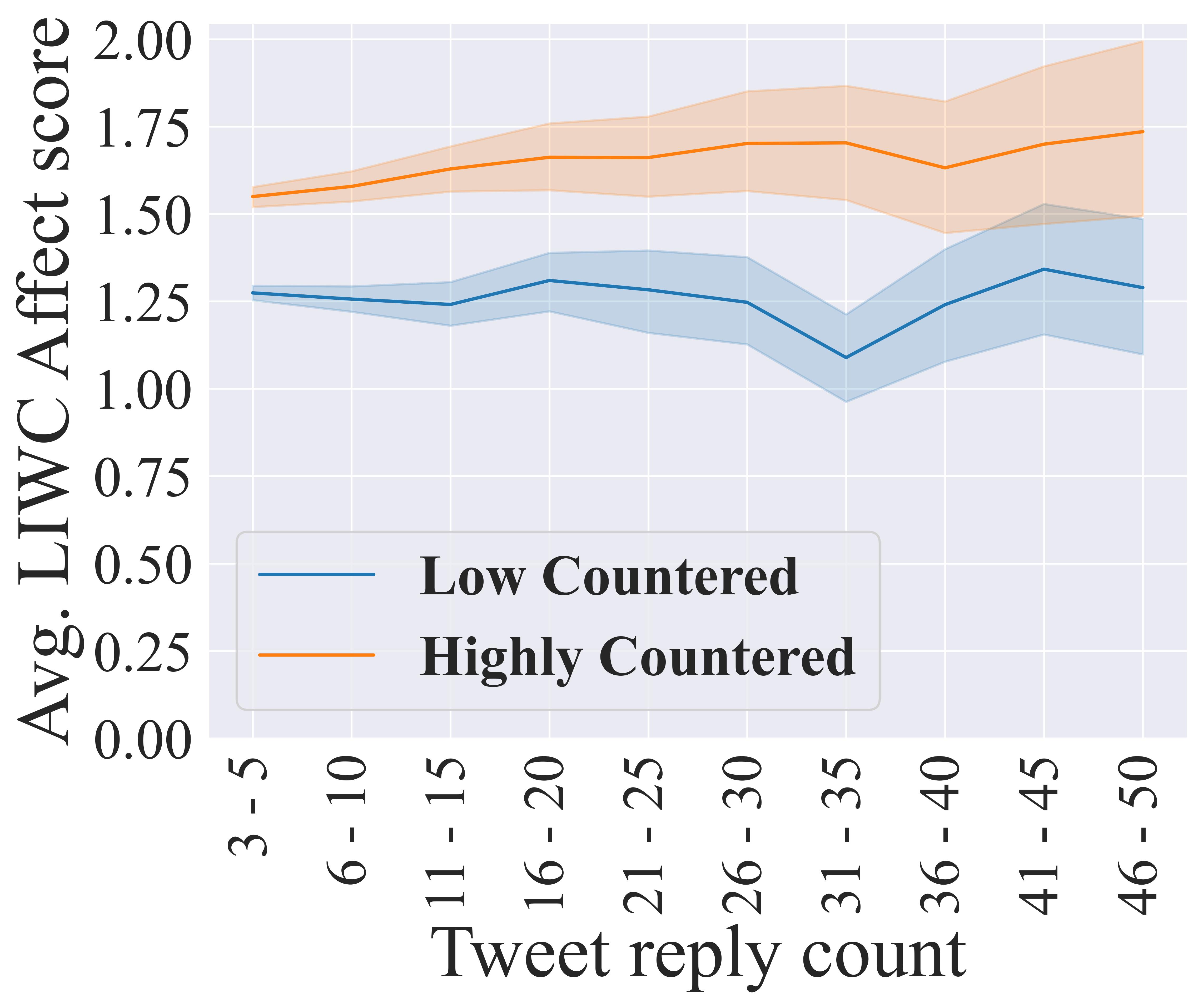}}
        \caption{LIWC Affect score}
        \label{subfig:liwc_affect}
    \end{subfigure}
    \hfill
    \begin{subfigure}[t]{0.24\textwidth}
        \raisebox{-\height}{\includegraphics[width=\textwidth]{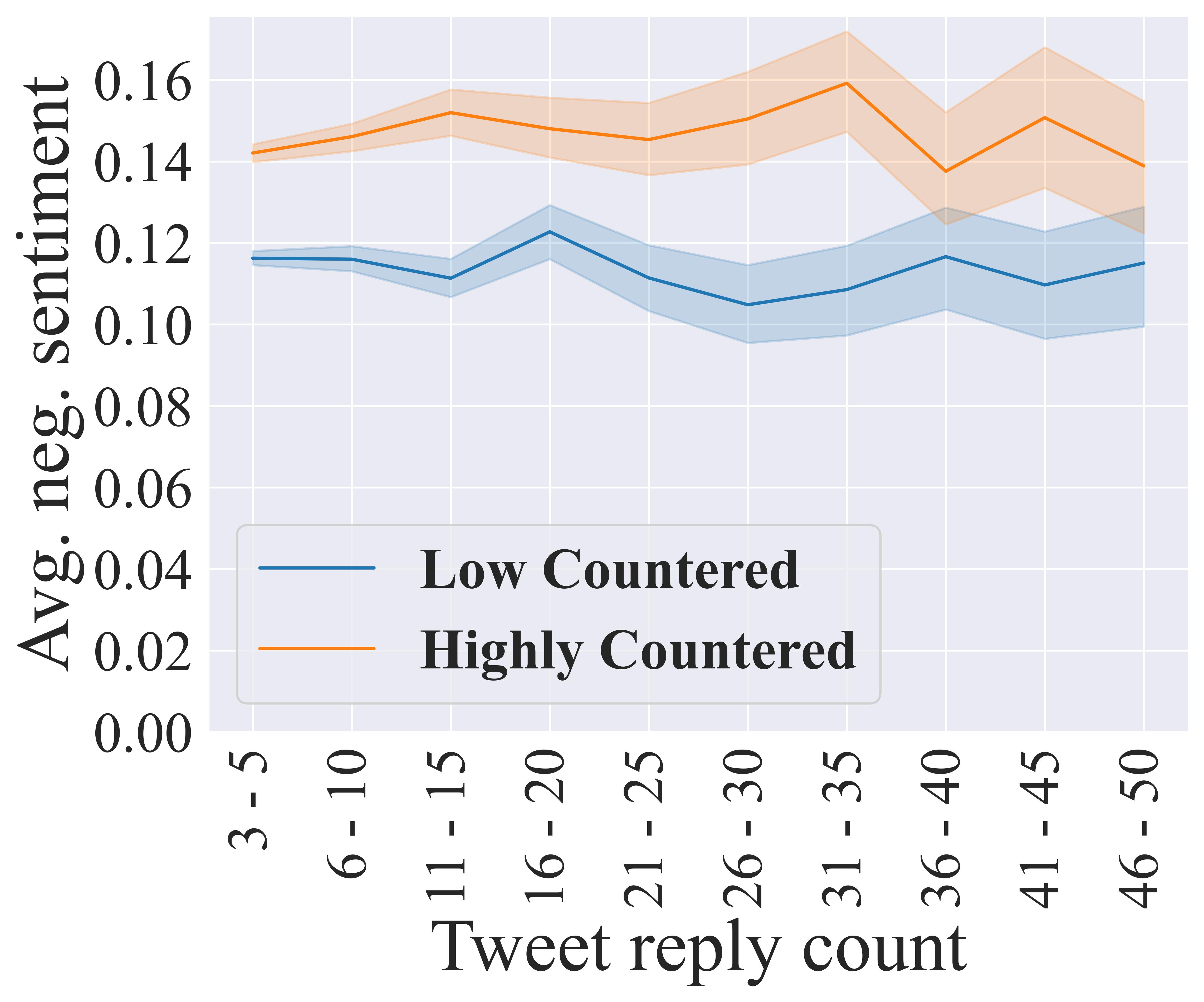}}
        \caption{VADER negative sentiment}
        \label{subfig:vader_neg}
    \end{subfigure}
    \hfill
    \begin{subfigure}[t]{0.24\textwidth}
        \raisebox{-\height}{\includegraphics[width=\textwidth]{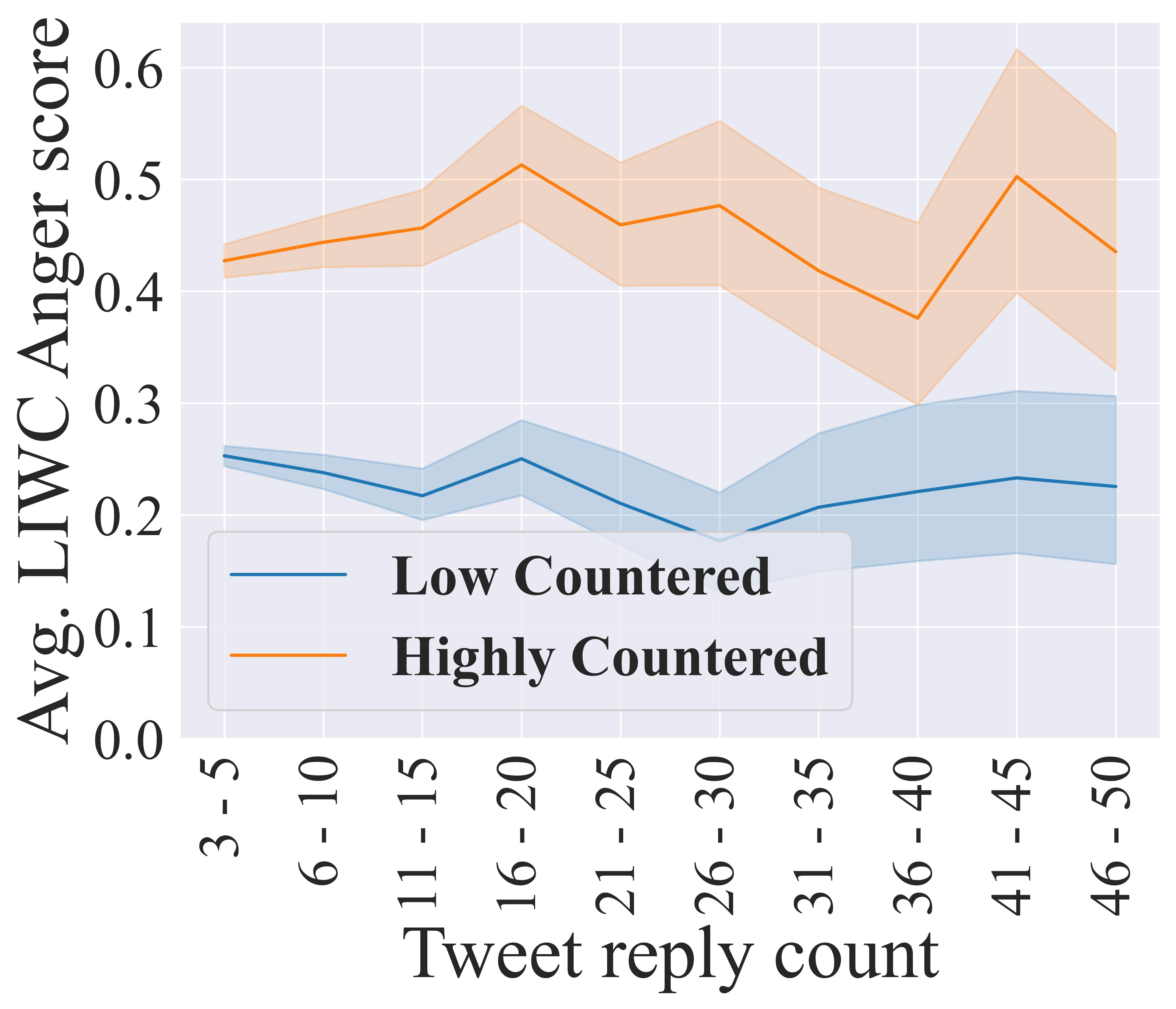}}
        \caption{LIWC Anger score}
        \label{subfig:liwc_anger}
    \end{subfigure}
    \hfill
    \begin{subfigure}[t]{0.24\textwidth}
        \raisebox{-\height}{\includegraphics[width=\textwidth]{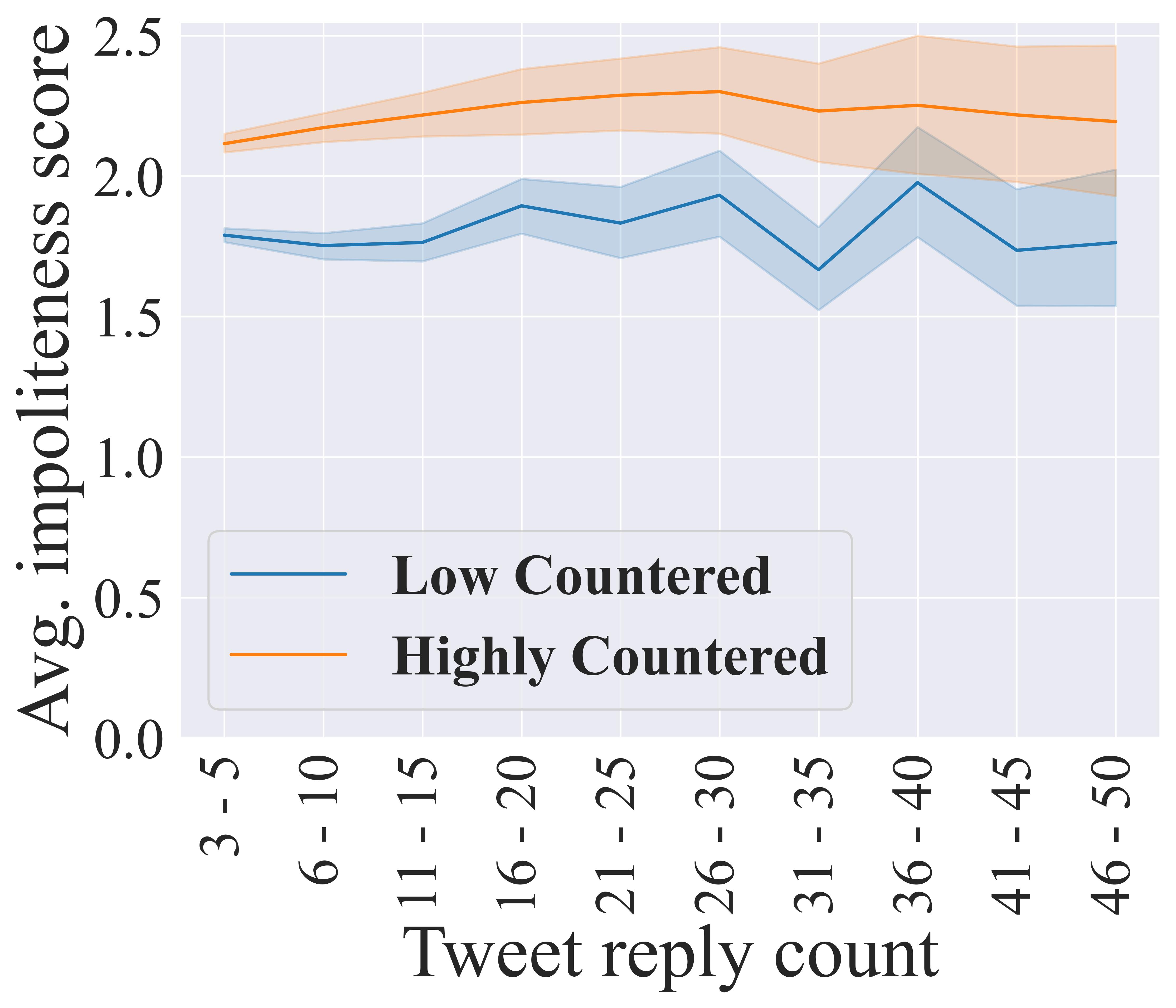}}
        \caption{Impoliteness score}
        \label{subfig:politeness}
    \end{subfigure}
    \hfill
    \begin{subfigure}[t]{0.24\textwidth}
        \raisebox{-\height}{\includegraphics[width=\textwidth]{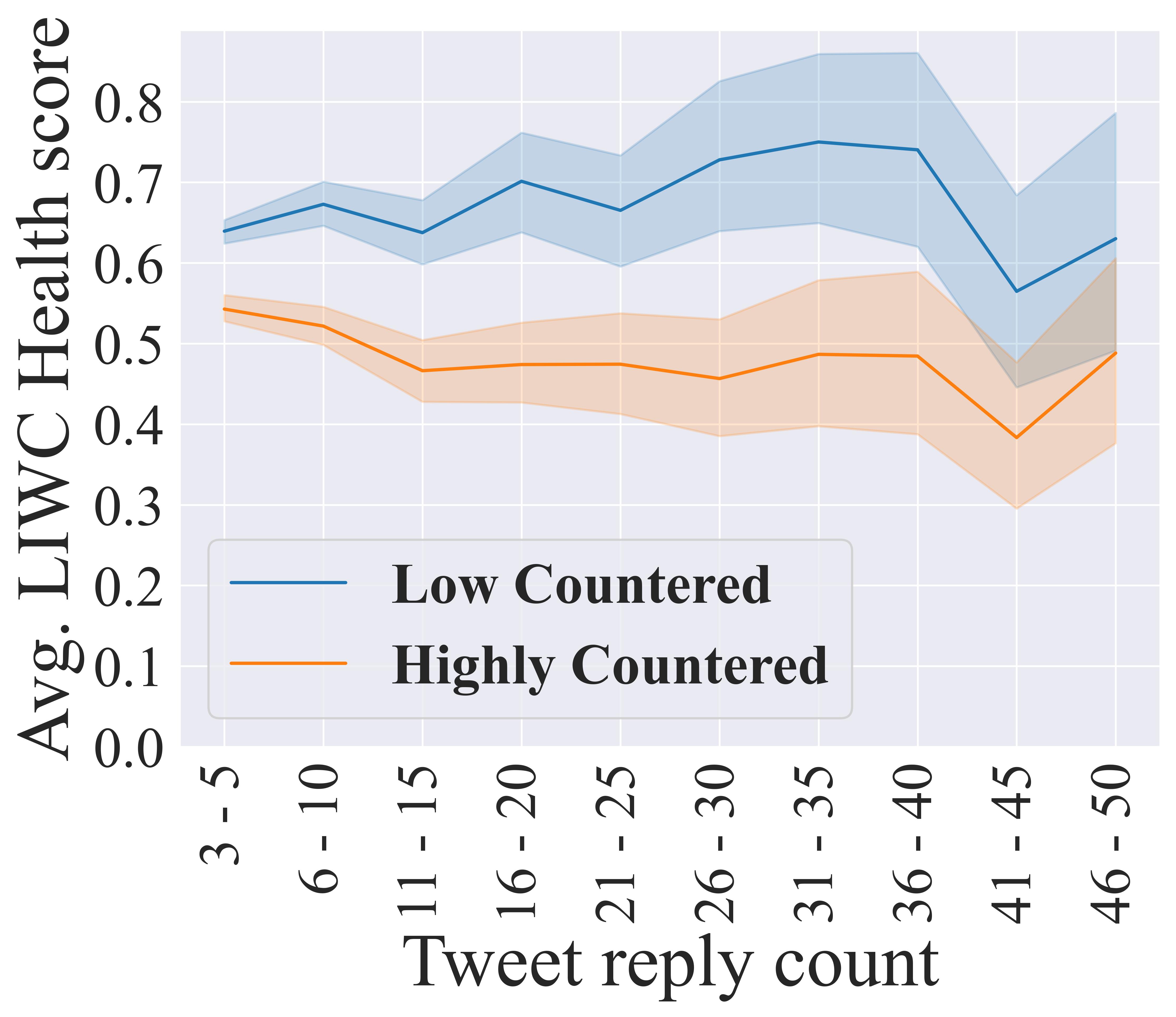}}
        \caption{LIWC Health score}
        \label{subfig:liwc_health}
    \end{subfigure}
    \hfill
    \begin{subfigure}[t]{0.24\textwidth}
        \raisebox{-\height}{\includegraphics[width=\textwidth]{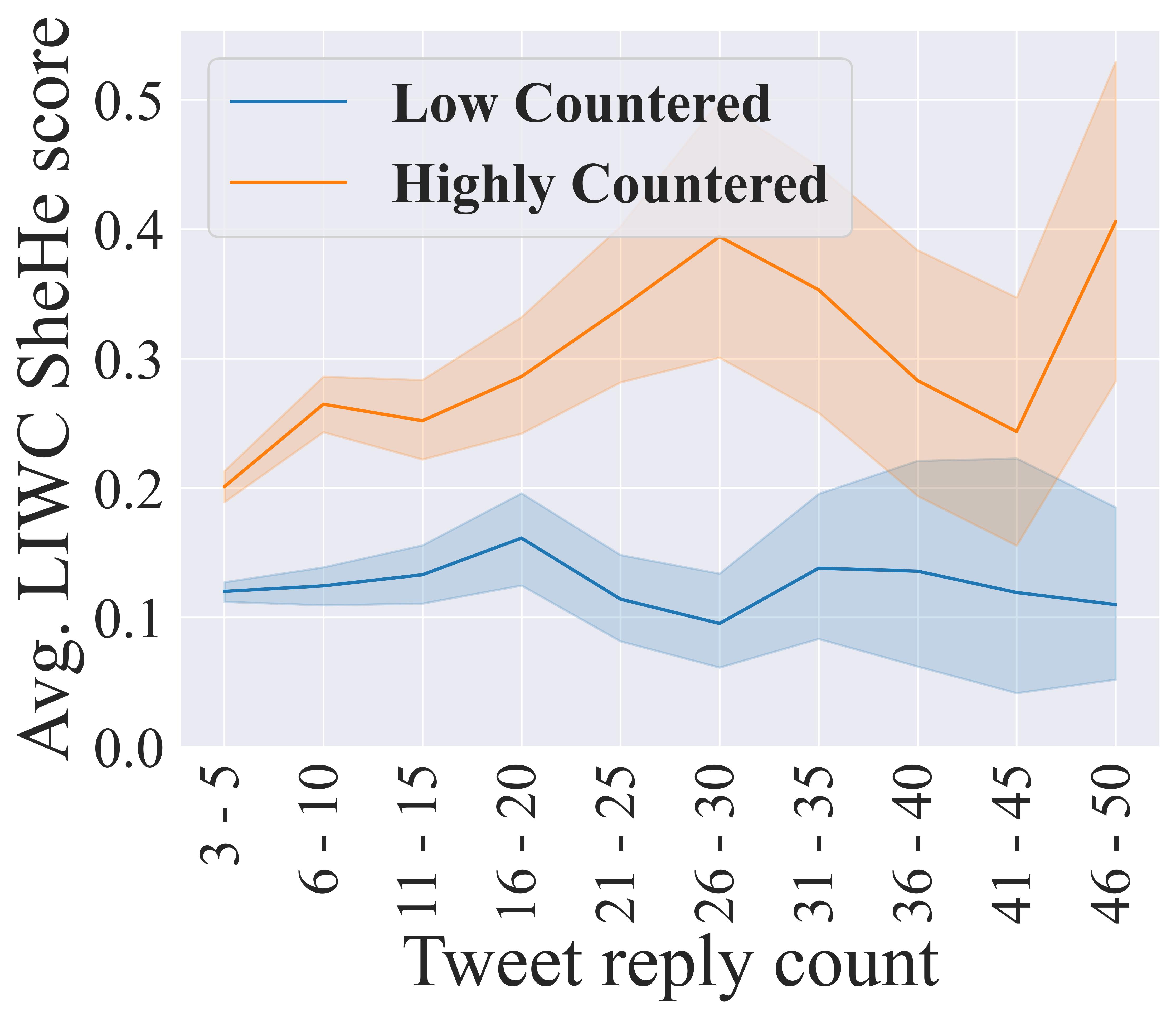}}
        \caption{LIWC SheHe score}
        \label{subfig:liwc_shehe}
    \end{subfigure}
    \hfill
    \begin{subfigure}[t]{0.24\textwidth}
        \raisebox{-\height}{\includegraphics[width=\textwidth]{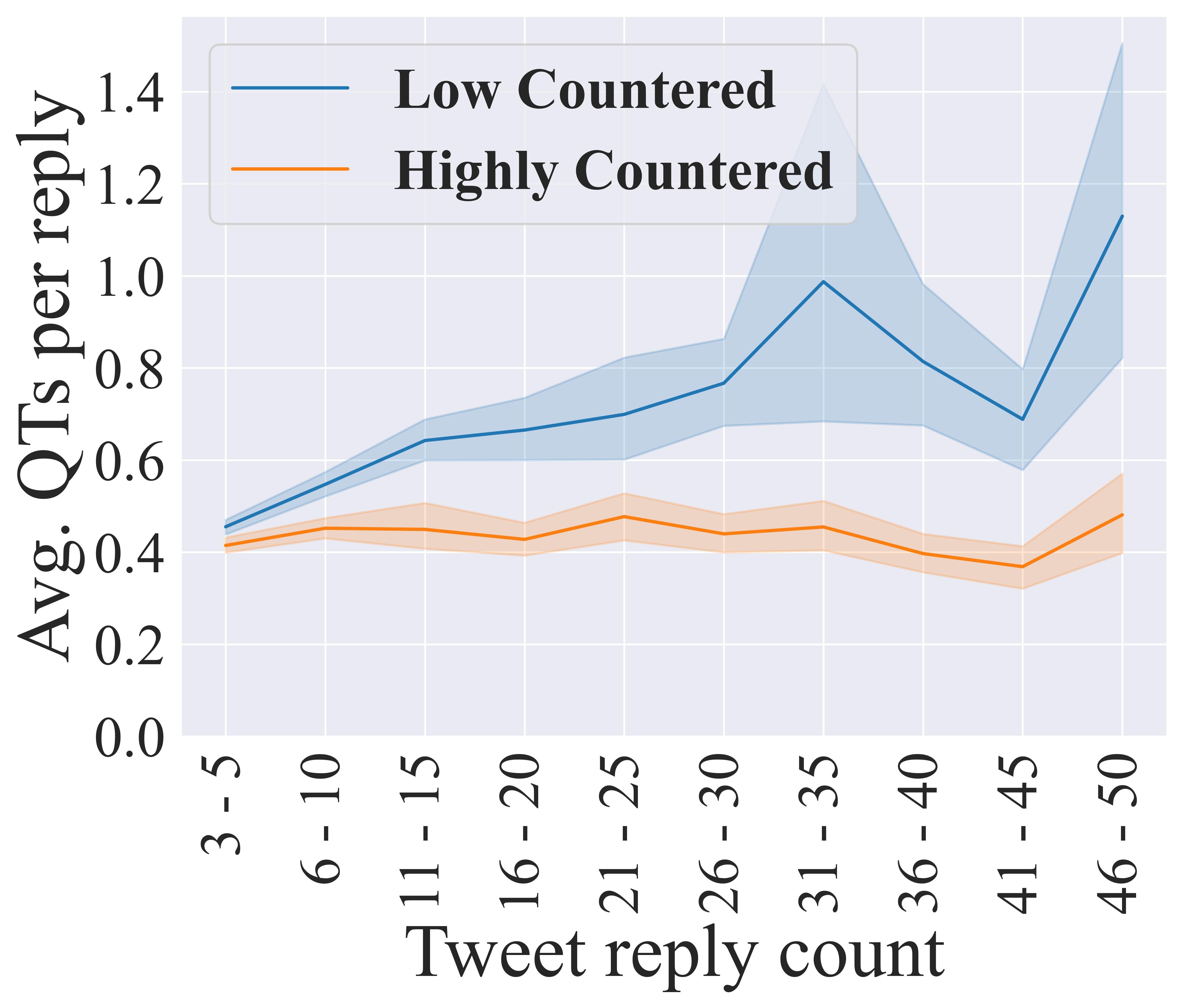}}
        \caption{Fraction of quote tweets (QTs) among all the replies}
        \label{subfig:scaled_qt}
    \end{subfigure}
    \hfill
    \begin{subfigure}[t]{0.24\textwidth}
        \raisebox{-\height}{\includegraphics[width=\textwidth]{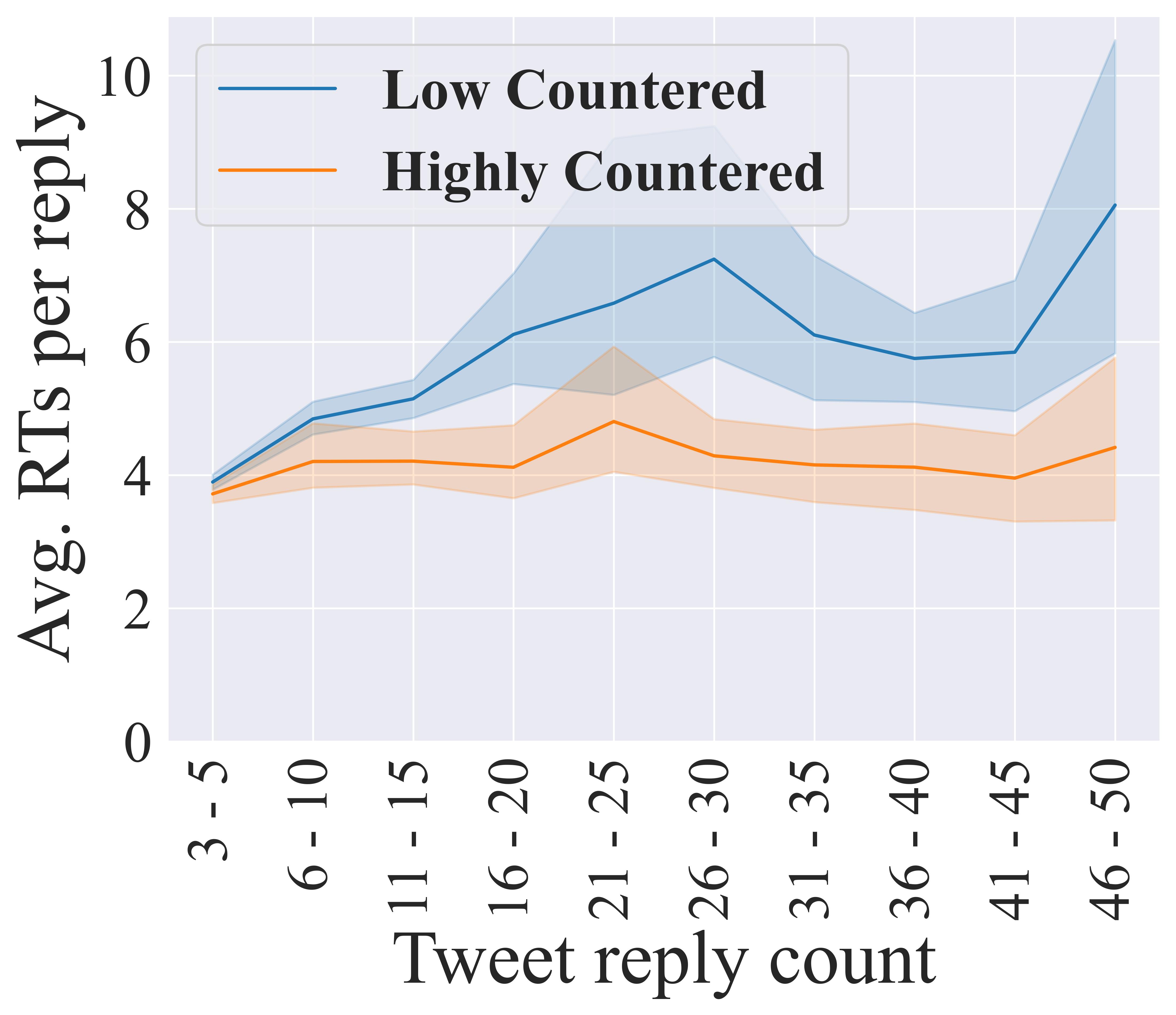}}
        \caption{Fraction of retweets (RTs) among all the replies}
        \label{subfig:scaled_rt}
    \end{subfigure}
    \caption{Means and 95\% confidence intervals of the linguistic and engagement attributes of misinformation tweets that get highly countered versus those that do not.}
    \label{fig:stratified_results}
\end{figure*}

\footnotetext{This statistical test was performed using Welch's unequal variances \textit{t}-test between the upper and lower quartiles (with respect to the proportion of counter-replies) of the data visualized in Figure~\ref{fig:disbelief_prop}.}

\subsection{Linguistic Attributes of Tweets that are Countered}\label{sec:feature_analysis_linguistic}

First, we observe from Figure \ref{subfig:liwc_affect} that on average, ``Highly Countered'' tweets contain 32.1\% higher usage of affective language (words and phrases that appeal more to emotions) than ``Low Countered'' tweets (p < 0.05 for all strata\footnote{all p-values in Sections ~\ref{sec:feature_analysis_linguistic},  ~\ref{sec:feature_analysis_engagement}, and ~\ref{sec:feature_analysis_user} are calculated using Welch's unequal variances \textit{t}-tests.}; average Cohen's d = 0.277)\footnote{``average Cohen's d'' here (and elsewhere in this paper) refers to the unweighted average of Cohen's d values of each stratum.}. This indicates that those who post counter-replies tend to gravitate more towards replying to misinformation that induces a stronger emotional reaction in them. This is consistent with the finding that emotional content gets more attention on social media in existing research works~\cite{schreiner2021impact}.
Further, we find that ``Highly Countered'' tweets express significantly higher negative sentiment than ``Low Countered'' tweets across all strata. Figure \ref{subfig:vader_neg} shows this result for VADER negative sentiment (p < 0.05 for all strata; average Cohen's d = 0.304); we find similar results for the ``negative emotion'' dimension of the LIWC lexicon (p < 0.05 for all strata; average Cohen's d = 0.279). In particular, we find that on average, ``Highly Countered'' tweets contain 104\% more anger-related words than ``Low Countered'' tweets (see Figure \ref{subfig:liwc_anger}) (p < 0.01 for all strata; average Cohen's d = 0.347). This implies that the negative tone of misinformation tweets attract more attention~\cite{huang2007attention, haselmayer2019fighting}, and therefore, more counter-replies. 

In addition, we measure differences in the degree to which the misinformation tweet expresses politeness and impoliteness. We do this by identifying the sets of linguistic strategies associated with each as presented in \cite{DanescuNiculescuMizil2013ACA}, and compute the total number of linguistic instances associated with each set to derive the ``politeness'' and ``impoliteness'' score, respectively. As shown in Figure \ref{subfig:politeness}, on average, ``Highly Countered'' tweets utilize 23.1\% more strategies associated with impoliteness than ``Low Countered'' tweets (p < 0.05 for all but one stratum; average Cohen's d = 0.248); this finding is consistent with the previous findings involving negative sentiment. Meanwhile, we do not find a significant difference between the groups for strategies associated with politeness, implying that trying to be polite in presenting a misinformation tweet does not significantly impact the chance of being countered.

Next, we find that there exist differences in topical presence between ``Highly Countered'' and ``Low Countered'' tweets. Figure~\ref{subfig:liwc_health} shows that on average, ``Highly Countered'' tweets utilize 28.7\% fewer health-related terms than ``Low Countered'' tweets (p < 0.05 for all but one stratum; average Cohen's d = 0.220). This suggests that for the average counter-reply poster, the inclusion of more technical medical terminology might pose a barrier for their willingness or ability to post an effective debunking response. One possible reason is that the inclusion of technical health-related terms can signal authority over the topic and be more convincing to the reader~\cite{buehl2001profiling, habernal2016makes}.

We also find that ``Highly Countered'' tweets use 2.5 times more third-person pronouns (e.g., `he', `she', `they', `them', etc.) than ``Low Countered'' tweets (p < 0.05 for all but one stratum; average Cohen's d = 0.259; see Figure~\ref{subfig:liwc_shehe}). 

\subsection{Engagement Attributes of Tweets that are Countered}\label{sec:feature_analysis_engagement}

In this subsection, we study the impact of engagement attributes (e.g., likes, retweets, etc.) on whether misinformation gets countered. There are two possibilities: (1) first, misinformation tweets with higher engagement get countered more often because the misinformation gets more attention and therefore, have a higher likelihood of becoming accessible to someone who would counter it; (2) second, misinformation tweets that get countered are less likely to be liked or retweeted by others. We investigate which of the two possibilities hold as per the data. 

In addition to the reply count, we compare tweets using the number of likes, retweets (RT), and quotes (QT) they receive. As these methods of engagement on the platform serve a different purpose and have different functionality than the ``reply'' method, it is worth using these metrics in our cross-group comparison. In order to effectively capture these differences with respect to reply count, we first perform a scaling of these attributes by dividing by the reply count, then performing comparisons of this quotient across the two groups.

Figure \ref{subfig:scaled_qt} shows that on average, ``Highly Countered'' tweets receive 37.6\% fewer QTs relative to replies on average (p < 0.05 for all strata). This difference is very small at the lowest stratum (8.9\% fewer; Cohen's d = 0.05), but is much higher on the highest stratum (57.4\% fewer; Cohen's d = 0.37). 
We receive similar results for retweets and likes; on average, ``Highly Countered'' tweets receive 27.4\% fewer retweets relative to replies (p < 0.05 for all but one stratum; see Figure \ref{subfig:scaled_rt}) and 25.6\% fewer likes relative to replies (p < 0.05 for all but 3 strata).

These findings show that the presence of counter-replies on a tweet organically decreases engagement by average users, suggesting that the practice of countering is potentially effective at reducing the spread of misinformation~\cite{colliander2019fake, wijenayake2020effect,friggeri2014rumor}.

\subsection{User Attributes of Tweet Posters that are Countered}\label{sec:feature_analysis_user}
First, we study the impact of the user being verified on Twitter on the tweet getting countered. We find that, on average, the proportion of ``Highly Countered''  misinformation posters that are verified is 16.8\% higher than that for ``Low Countered'' misinformation posters (p < 0.05 in all but 3 strata; average Cohen's d = 0.143). 

Since the majority of the posters on Twitter are non-verified, we study that set of users next. We compare the attributes of non-verified users in the ``Highly Countered'' group versus the ``Low Countered'' group. 
For the remainder of the attributes, we found none of them to be statistically different across the two groups. 
Thus, together with the linguistic results presented in Section \ref{sec:feature_analysis_linguistic}, we find that the content of the misinformation tweet is more important in attracting countering than the user who posts the misinformation.

\section{Inequality In Social Correction}\label{sec:ineq}
We further investigate the potential inequality in social correction. This can help identify whether certain types of users are less likely to be countered, leading to an increase in disparity. 
Motivated by existing work~\cite{verma2022examining}, we use education level as a key demographic variable to illustrate the potential inequality between different users. Since lack of education and literacy play a crucial role in believing in misinformation~\cite{roozenbeek2020susceptibility, georgiou2020covid, van2017education}, it is important to study whether it also impacts correction.

We derived the education level of users by quantifying the readability of their posts using the Automated Readability Index (ARI), which is known to produce an approximate representation of education level in prior works~\cite{senter1967automated, flekova2016exploring, rajadesingan2015sarcasm}. A higher ARI corresponds to a higher education level.
We use the ``pre-misinformation'' posts of each user (i.e., posts made within the 7 days prior to posting the misinformation tweet) to calculate that user's ARI. 
Then, for each post, we compute the ARI score~\cite{senter1967automated, flekova2016exploring, rajadesingan2015sarcasm}. Finally, we compute the average of these scores, and use it as the final ARI value to present the education level of the user.
Thus, it should be noted that the ARI score is \textit{not} the education level portrayed in the misinformation tweet, but instead, the education level derived across the \textit{historical} posts of the user who spread misinformation tweets. We randomly sampled 10,000 users who spread misinformation in our dataset to illustrate the inequality phenomenon.

\begin{figure}[!htbp]
    \centering
    \includegraphics[width=0.42\textwidth]{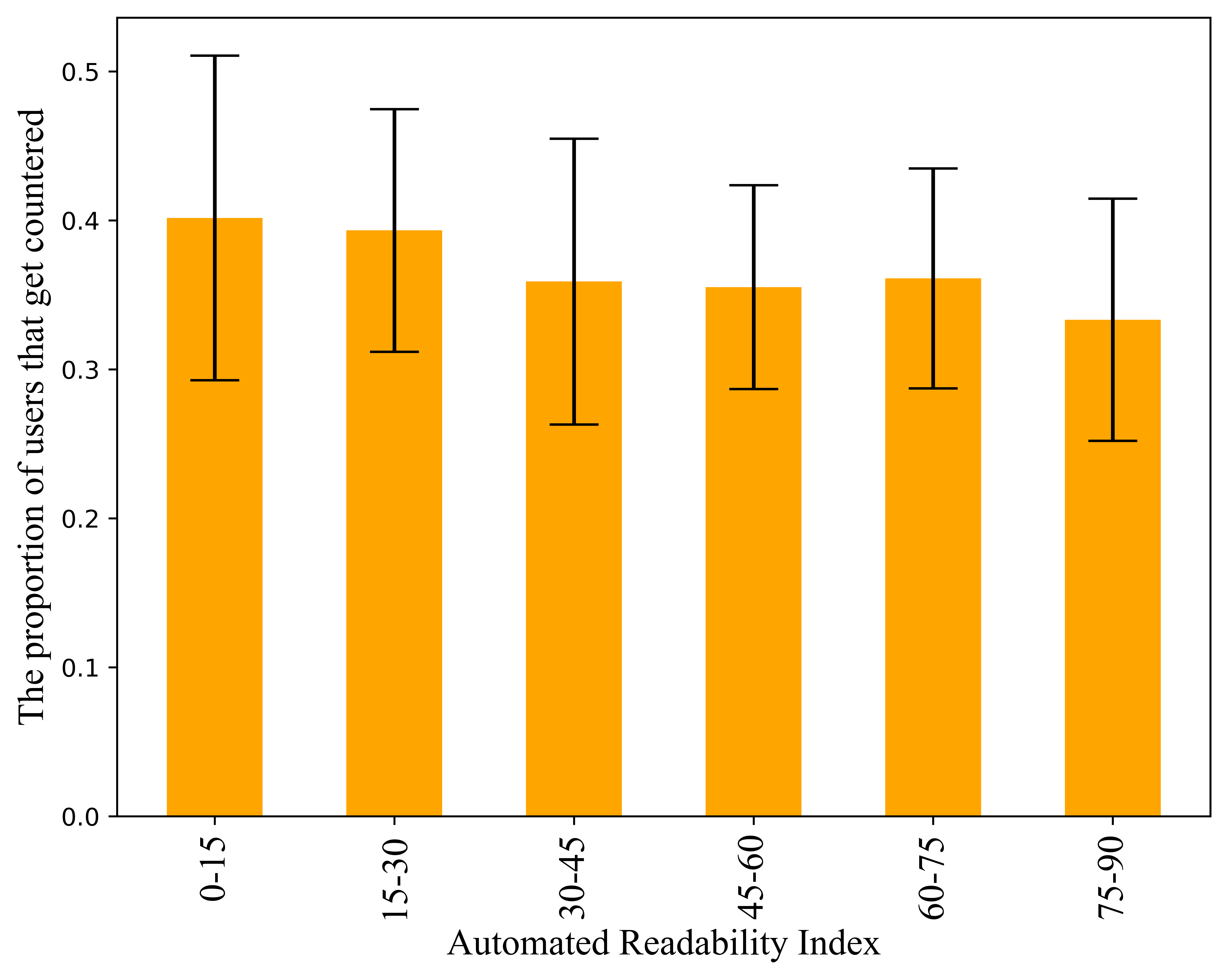}
    \caption{Comparison of user communities with different education levels. As shown, users with lower education levels will have higher possibilities of getting countered when sending misinformation tweets. 
    }
    \label{fig:education_demo}
\end{figure}

As shown in Figure~\ref{fig:education_demo}, we find that misinformation posts made by users with lower education levels have a higher likelihood of getting corrected. 
There is a systematically negative trend with an increase in the user's (perceived) education level. This highlights a need to pay attention to misinformation spread by users who portray a higher education level, since ordinary users are less likely to correct them.

\section{Social Correction Prediction}\label{sec:prediction}

In this section, we aim to answer two research questions:
\begin{itemize}
    \item RQ1: Given a misinformation tweet, can we predict whether it will be countered or not in the future?
    \item RQ2: Given a misinformation tweet that will be countered in the future, can we predict whether it will be countered with fewer or more counter-replies?
\end{itemize}

Both RQs are important to address for the combating of future misinformation. By being able to effectively predict future interactions surrounding misinformation tweets, we can better identify sets of online interactions where misinformation is being organically countered, along with those where additional countering needs to be performed. Answering RQ1 can identify sets of misinformation posts where other users may take the initiative in posting a counter-reply, while answering RQ2 can predict the intensity or magnitude of countering. 

\subsection{Dataset}
For both the research questions, we use the aforementioned dataset as we described in Section~\ref{sec:feature_analysis}. 

For RQ1, we divide the dataset into two sets of misinformation tweets: (1) misinformation tweets that have replies but none of them are counter-replies; (2) misinformation tweets that have at least one counter-reply. The sizes of these sets are 17,787 and 55,136, respectively.
For RQ2, we divide the misinformation tweets into two groups: one with a low proportion of counter-replies, and another group with a high proportion of counter-replies. Similar to the stratified setup in Section~\ref{sec:feature_analysis_strat}, we use the proportion of counter-replies as an indicator of membership for the two groups. 
The bottom 25\% of posts with respect to their countering proportion are assigned to the low countered group. On the other hand, the top 25\% posts with the highest proportion of countering replies are assigned to the highly countered group. The sizes of these sets are 14,274 and 15,224, respectively.

\subsection{Experimental setup}

Using each of these datasets, we follow similar approaches in tweet prediction tasks~\cite{micallef2020role, zahera2019fine} to address both RQ1 and RQ2.
We aim to build a binary classifier for each of RQ1 and RQ2, using the label definitions described above. For both RQs, we use the same set of features. We begin with the set of attributes listed in Table~\ref{tab:attr_list} with p < 0.001 and have non-null values for all datapoints; there are 63 such attributes (53 linguistic, 5 engagement, 5 poster). As shown in the existing tweet prediction task~\cite{micallef2020role}, the semantic information from textual embedding benefits the prediction task. Thus, we also generate the embedding vector for each tweet using RoBERTa~\cite{liu2019roberta}, which results in a 768-dimensional feature vector. Finally, we concatenate the above feature vectors to form a tweet feature vector to comprehensively represent the tweet and use it for classification.

\textbf{Classifier}: 
Following similar tweet or general text classification tasks~\cite{micallef2020role, he2021racism, he2021petgen}, 
we deploy widely-used conventional machine learning classifiers including Logistic Regression, XGBoost, and a Feed-forward Neural Network with a single hidden layer, using the feature vector as input. 
During the experiment, 10-fold cross-validation is deployed, and we report precision, recall, and F-1 score as the performance metrics. 

\subsection{Classifier Performance}

\begin{table}[!tbp]
    \centering
    \begin{tabular}{lrrr}
    \hline
    Method &  Precision &  Recall &  F-1 score \\ \hline
    Logistic Regression          &      0.801 &   0.929 &     0.860 \\
    XGBoost &      0.803 &   0.908 &     0.852 \\
    Neural Network                &      0.804 &   0.914 &     0.855 \\ \hline
    \end{tabular}
    \caption{RQ1: Classifier performance of whether tweets will get countered or not. }
    \vspace{-15pt}
    \label{tab:result_rq1}
\end{table}

In Table~\ref{tab:result_rq1}, we report the classification result for RQ1. As we can see, all three models are able to achieve good performance on the task. The logistic regression achieves the best performance in terms of precision, recall, and F-1 score; this result is also found in other similar tweet classification tasks~\cite{micallef2020role}. This high performance grants the ability to effectively predict whether a tweet will be countered or not, enabling fact-checkers and social media platforms to prioritize countering tweets identified as less likely to be countered organically.

\begin{table}[!tbp]
    \centering
    \begin{tabular}{lrrr}
    \hline    
    Method &  Precision &  Recall &  F1 score \\ \hline
    Logistic Regression         &      0.731 &   0.742 &     0.737 \\
    XGBoost             &      0.841 &   0.756 &     0.796 \\
    Neural Network                &      0.848 &   0.759 &     0.801 \\ \hline
    \end{tabular}
    \caption{RQ2: Classification performance of whether tweets will be highly countered versus that will be low countered.}
    \vspace{-15pt}
    \label{tab:result_rq2}
\end{table}

For RQ2, the classification result is shown in Table~\ref{tab:result_rq2}. As we can see, the model performance is still reasonably acceptable, but is worse compared to RQ1. 
This decrease in performance may imply that the task to identify the \textit{intensity} of countering tweets is not only more difficult, but also distinct from the task to identify \textit{whether} a tweet will be countered. In other words, the phenomenon of posting of the first counter-reply is easier to forecast than that of the posting of additional counter-replies given that at least one has already been posted.

\section{Discussion and Conclusion}

In this paper, we studied the tweet and user-level properties of misinformation tweets that get countered versus those that do not. The in-depth analysis shows that misinformation tweets expressing negative emotion, strong emotion, third-person pronouns, and strategies associated with impoliteness are more likely to result in more countering replies from users. Our result also shows that tweets that get countered have a higher amount of reply engagement in proportion to like, retweet, and quote tweet engagement. Moreover, we develop well-performing classifiers to predict whether a misinformation tweet will be countered or not, and if so, to what degree they will be countered (i.e. the proportion of its replies that end up being counter-replies).

Given the statistical significance of our analysis and the high performance of our classifiers, we demonstrate that it is possible to identify tweets that are more or less likely to get countered. In particular, nearly all of these attributes (tweet linguistic attributes and user attributes) are readily available as soon as the tweet is posted, allowing for the quantity of future counter-misinformation (or the lack thereof) to be reasonably forecast. This can have major implications in times of breaking news or other such events in which large quantities of (mis-)information are posted to online platforms at a rapid rate; in conjunction with state-of-the-art misinformation detection approaches, the counter-reply prediction approach presented in this paper can be used to identify tweets that are less likely to be countered, possibly necessitating additional platform-level approaches to control the spread of misinformation for these tweets. One of these approaches may be adding or increasing interventions to draw attention towards accuracy, an approach that has been shown to be effective in discouraging users from spreading misinformation~\cite{pennycook2020}.

A limitation of this work that it focuses on only one platform: Twitter. On other online platforms, different mechanisms of post and user engagement, as well as information exchange, may be present~\cite{micallef2022cross}, possibly influencing the types of misinformation tweets and posters users will choose to counter. Another limitation is that it studies only one topic (COVID-19 vaccines), which has become one of the most widely discussed topics in our society due to the universal effects of the COVID-19 pandemic. On misinformation-related topics that might be more obscure or less widely discussed (e.g. flat earth theories), it could be possible that the more specific demographics of misinformation and/or counter-reply posters may affect the ways in which they interact. In addition, we only study text in the English language; the dynamics and discussion in other languages and other modalities (images, videos) may differ~\cite{verma2022overcoming}.

For future work, similar analysis can be performed on the user network surrounding the misinformation poster and counter-reply poster (e.g. their followers and those they follow, how much misinformation these accounts spread, etc.) in order to assess if there are any network-related attributes that may increase the likelihood of counter-replies. In addition, given that we can reasonably determine which tweets will and will not be countered, it would also be valuable to perform user studies or field studies to evaluate if certain characteristics about online encounters with misinformation can increase (or decrease) the likelihood of a user posting a counter-reply. Also, while we explore it in Section~\ref{sec:ineq}, further studies can be done to understand the inequities surrounding counter-reply targets along additional demographic, social, political, and/or geographic dimensions; this can allow further exploration of the greater societal implications surrounding counter-misinformation.

\noindent \textbf{{ACKNOWLEDGMENTS}} 
This research/material is based upon work supported in part by 
NSF grants CNS-2154118, IIS-2027689, ITE-2137724, ITE-2230692, CNS-2239879, and funding from Microsoft, Google, and Adobe Inc. Any opinions, findings and conclusions or recommendations expressed in this material are those of the author(s) and do not necessarily reflect the position or policy of 
NSF and no official endorsement should be inferred. 
We thank the CLAWS research group members for their help on the project.


\bibliographystyle{ACM-Reference-Format}
\balance

\bibliography{main}

\appendix

\end{document}